\RequirePackage{fix-cm}

\documentclass{svjour3}
\smartqed  % flush right qed marks, e.g. at end of proof

\usepackage{graphicx}

\usepackage{microtype}
\usepackage{url}
\usepackage{amsmath}
\usepackage{listings}
\usepackage{color}
\usepackage{amssymb}
\usepackage{wrapfig}
\usepackage{algorithm}
\usepackage[noend]{algpseudocode}
\algrenewcommand\algorithmicindent{0.8em}
\usepackage{cite}
\usepackage{subcaption}
\usepackage{rotating}
\usepackage{hyperref}

\usepackage{xcolor}
\hypersetup{
    colorlinks,
	linkcolor={blue}, %{red!50!black},
	citecolor={blue}, %{blue!50!black},
	urlcolor={blue}   %{blue!80!black}
}

\newsavebox{\imagebox}

\journalname{}

\begin{document}

\title{OODIDA: On-board/Off-board Distributed Real-Time Data Analytics for Connected Vehicles\thanks{The final authenticated version is available online at \url{https://doi.org/10.1007/s41019-021-00152-6}.}}

\titlerunning{OODIDA: On-board/Off-board Distributed Data Analytics}        % if too long for running head

\author{Gregor Ulm
       \and Simon Smith
       \and Adrian Nilsson
       \and Emil Gustavsson
       \and Mats Jirstrand
}

\authorrunning{G.~Ulm~et~al.} % if too long for running head

\institute{Gregor Ulm \at
           Fraunhofer-Chalmers Research Centre for Industrial Mathematics \\
           and Fraunhofer Center for Machine Learning\\
           Chalmers Science Park\\
           SE-412 88 Gothenburg, Sweden\\
           \email{gregor.ulm@fcc.chalmers.se}
           \and
           Simon Smith \at
           Fraunhofer-Chalmers Research Centre for Industrial Mathematics \\
           and Fraunhofer Center for Machine Learning\\
           Chalmers Science Park\\
           SE-412 88 Gothenburg, Sweden\\
           \email{simon.smith@fcc.chalmers.se} 
           \and
           Adrian Nilsson \at
           Fraunhofer-Chalmers Research Centre for Industrial Mathematics \\
           and Fraunhofer Center for Machine Learning\\
           Chalmers Science Park\\
           SE-412 88 Gothenburg, Sweden\\
           \email{adrian.nilsson@fcc.chalmers.se} 
           \and
           Emil Gustavsson \at
           Fraunhofer-Chalmers Research Centre for Industrial Mathematics \\
           and Fraunhofer Center for Machine Learning\\
           Chalmers Science Park\\
           SE-412 88 Gothenburg, Sweden\\
           \email{emil.gustavsson@fcc.chalmers.se} 
           \and
           Mats Jirstrand \at
           Fraunhofer-Chalmers Research Centre for Industrial Mathematics \\
           and Fraunhofer Center for Machine Learning\\
           Chalmers Science Park\\
           SE-412 88 Gothenburg, Sweden\\
           \email{mats.jirstrand@fcc.chalmers.se} 
}

%\date{Received: date / Accepted: date}
% The correct dates will be entered by the editor

\maketitle

\begin{abstract}
A fleet of connected vehicles easily produces many gigabytes of data per hour, making centralized (off-board) data processing impractical. In addition, there is the issue of distributing tasks to on-board units in vehicles and processing them efficiently. Our solution to this problem is OODIDA (On-board/Off-board Distributed Data Analytics), which is a platform that tackles both task distribution to connected vehicles as well as concurrent execution of tasks on arbitrary subsets of edge clients. Its message-passing infrastructure has been implemented in Erlang/OTP, while the end points use a language-independent JSON interface. Computations can be carried out in arbitrary programming languages. The message-passing infrastructure of OODIDA is highly scalable, facilitating the execution of large numbers of concurrent tasks.

\end{abstract}

\section{Introduction}
Big data in the automotive industry is of increasing concern, considering that connected vehicles may produce large volumes of data per hour. When dealing with a fleet of vehicles, centrally processing such data is impractical, if not infeasible. However, with OODIDA (On-board/Off-board Distributed Data Analytics), which is a platform that facilitates the distribution and concurrent execution of real-time data analytics tasks in a heterogeneous system, we can conveniently process vehicle telemetry data as batches or pseudo-realtime streams close to the data source. This is a step towards facilitating efficient big data analytics in vehicular networks. The majority of the computational work is carried out on client devices (on-board), so-called edge devices, and only a supplementary part on the server (off-board).

\begin{figure}[h]
\centering
\includegraphics[scale=1.0, trim=6.9cm 14.9cm 25.4cm 3.5cm, clip]{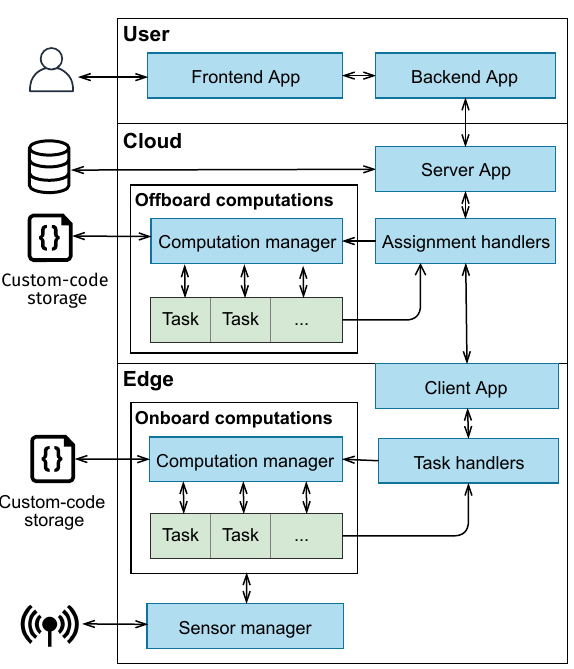}
\caption{The modular architecture of OODIDA. The backend, server, and client apps as well as the assignment and task handlers have been implemented in Erlang. The user frontend, computation manager and sensor manager have been implemented in Python. Modules or functions that carry out tasks can be implemented in an arbitrary programming language. The sensor manager (bottom) interfaces with the vehicle's CAN bus.}
\label{fig:architecture}
\end{figure}

OODIDA uses a virtual private network for communication. It connects data analysts with a large number of on-board units (OBUs). While our system could be used for general distributed data processing tasks, it has been designed for data exploration and rapid prototyping in the automotive domain, targeting a fleet of reference vehicles. An architecture diagram is provided in Fig.~\ref{fig:architecture}: $m$ users interact with a central cloud application that is connected with $n$ client nodes. The main feature of our system is distributing and concurrently processing decentralized data analytics tasks of arbitrary duration. We provide a JSON interface for external applications. In addition, error handling mechanisms address the issue of intermittent availability of client devices.

This paper presents the architecture and design of OODIDA as well as use cases that illustrate how this framework solves real-world problems. It has been deployed to test vehicles and used for processing live real-time data. Due to legal agreements that prevent us from publicly sharing source code artifacts, we need to keep some details under wraps. That being said, we do share implementation details of the server in Sect.~\ref{sol:bridge}. This is the central part of our system, which facilitates the high level of responsiveness and ability to concurrently process large numbers of tasks. The provided description of our system makes it possible to retrace our work. We thus believe that the information we share is sufficient to facilitate the creation of a similar system, which should largely offset that we cannot publicly release the source code of the software we created.

From a conceptual perspective, OODIDA is noteworthy for applying the paradigm of lightweight concurrent processing, via the programming language Erlang, to the automotive domain for real-time data analytics. This makes it possible to execute concurrent workloads. The system is also scalable as we sidestep issues related to threaded execution. Our approach is also less resource-heavy than a containerized solution where multiple containers are executed on the client.

The remainder of this paper starts off with relevant background information in Sect.~\ref{problem}. Subsequently, we provide a detailed description of OODIDA in Sect.~\ref{solution}, with a particular focus on how this system handles distributed concurrent execution of multiple assignments. This section also includes examples of how our general-purpose system for distributed real-time data analytics has been applied to various problems in the automotive domain. We substantiate the practical usefulness of our system with an empirical evaluation in Sect.~\ref{sec:eval} that focuses on scalability. The paper concludes with an overview of related work in Sect.~\ref{related}, which discusses various alternatives and why we did not use them, and potential future work in Sect.~\ref{future}.

\section{Background}
\label{problem}
In this section, we provide some background information on automotive big data. First, we summarize challenges commonly associated with big data, covering volume, velocity, and variety of data but also the issue of data privacy~(Sect.~\ref{p1}). Second, we look at examples of decentralized data processing, covering real-world use cases in our domain~(Sect.~\ref{p2}). Third, we highlight how scalability, based on transparent distribution and concurrent execution on client devices and the server, helps us solve automotive big data problems at the fleet level~(Sect.~\ref{p3}). Fourth, we clarify relevant terms~(Sect.~\ref{p4}).

\subsection{Big Data Challenges}
\label{p1}
The most immediate issue with data generated by connected vehicles is its immense \emph{volume} and \emph{velocity}, potentially exceeding 20 gigabytes per hour and vehicle~\cite{coppola2016connected}. It is impractical to continually transfer such volumes of data to a central server for processing, particularly at the fleet level. Communication protocols such as 5G and beyond will increase available bandwith. However, data generation outpaces those increases, and vehicular networks cannot realistically be expected to be stable. Thus, processing data on the edge and sending results in certain intervals is a more stable approach than continually sending data, even amounts that could conveniently be transferred via the network, to a central server for processing. With time-critical information, transferring data to a central server for subsequent processing may not be viable as the overhead of data transmission negatively affects processing times compared to processing data close to its source. There is also the issue of the \emph{variety} of big data due to collecting a multitude of signals. Thus, the heterogeneity of data rules out a one-size-fits-all approach. An additional challenge, but one that is externally imposed, is \emph{data privacy}~\cite{chen2012data, tene2011privacy}. A recent example is the General Data Protection Regulation (GDPR) of the European Union~\cite{gdpr}, which places heavy restrictions on data handling. On-board processing sidesteps data privacy issues that would have to be addressed with central data collection.

\subsection{Decentralized Big Data Processing}
\label{p2}
The main purpose of OODIDA is to make automotive big data easier to handle by processing it closer to its source, with the goal of limiting the amount of data that needs to be processed on a central server. To illustrate this idea, we present categories of practical use cases.

We start with use cases where the central server only collects results from clients, without further off-board processing. A straightforward example is \emph{filtering} on the client, e.g.\ monitoring or anomaly detection. Retaining only values that fulfill a given predicate may allow us to discard most data. A similar case is \emph{sampling}, where a randomly chosen subset of data is retained. However, such tasks may require additional processing. A related example is \emph{MapReduce}~\cite{dean2008mapreduce}. Consider the standard MapReduce problem of determining word frequency: clients first map each word $w$ of the input to the tuple $(w, 1)$. Afterwards, they reduce all those tuples to their respective count~$c$, i.e.\ $(w, c)$. This is the data that is sent to the server, which, in turn, reduces all incoming data to the final value $(w, c')$, where $c'$ specifies the total count of $w$ in the input. Large-scale MapReduce is hardly the goal of OODIDA, but some of its use cases fit this paradigm very well~(cf. Sect.~\ref{sec:usecases}).

A more complex use case is federated learning~\cite{mcmahan2016federated}, which is an example of distributed machine learning. In this case, the server maintains a global machine learning model, which is sent to clients. Each client trains the model with local data and subsequently sends the updated model to the server, which computes an average of the received updated local models. In practice, there is likely a limited number of iterations, based on the total error of the global model. Once it falls below a specified threshold value, training is considered complete. Subsequently, the final global model is sent to the user. Federated learning is an active area of research, with a particular interest from the mathematical optimization community. Examples include work on general distributed optimization~\cite{konevcny2016federated} and the alternating direction of multipliers (ADMM)~\cite{li2017robust}. Relevant for our case is that large-scale optimization on edge devices using deep federated learning has been shown as feasible~\cite{hardy2017distributed}.

Data processing tasks could be one-off with a fixed duration or they may run indefinitely long. In some scenarios, it makes sense to repeatedly perform narrowly defined tasks, for instance getting a status update every $n$ seconds, which can be modeled as an assignment with multiple iterations. Federated learning is merely a more complicated example of this approach as the results of iteration $i$ are used as input of iteration $i+1$. Another useful addition is the emulation of stream processing by iteratively processing batches of data. The shorter the iterations are, the closer we get to pseudo-realtime stream processing.

\subsection{Scalability at the Fleet Level}
\label{p3}
In order to handle big data problems at the fleet level, we need an effective means of task distribution. Of course, we also have to have the ability to issue multiple assignments to overlapping subsets of clients. This necessitates that client devices are able to concurrently execute tasks. Meanwhile, the central server has to remain responsive even as the workload on the system increases. That being said, there is a clear limitation to the amount of work the system needs to perform as we are not targeting a large fleet of production vehicles. Instead, the goal is to execute an analytics platform on a private cloud that connects to OBUs in test vehicles, so-called \emph{reference vehicle computational nodes}~(RVCNs). This facilitates rapid prototyping of data analytics methods which may eventually be executed on OBUs in production vehicles.

\subsection{Some Terminology}
\label{p4}
As this paper does not exclusively address a computer science audience, we would like to clarify a few relevant terms. The \emph{actor model} is a mathematical model for describing concurrent computations, developed by Hewitt~\cite{hewitt1973session}. In it, independent actors send and receive messages, making it straightforward to model concurrent computations. The most prominent language related to the actor model is Erlang, in which we have implemented the message-passing infrastructure of OODIDA. There is some conceptual confusion surrounding the terms \emph{concurrent} and \emph{parallel}, however. While even computer scientists may use them interchangeably, we follow, for instance, Harper~\cite{harper2016practical} and Marlow~\cite{marlow2012parallel}, and understand concurrency as the simultaneous execution of nondeterministic computations and parallelism as the simultaneous execution of deterministic computations. An example of the former is lock-free garbage collection; one of the latter is matrix multiplication. As we are not going to share Erlang source code artifacts, there is no need to explain syntactic details. However, we do frequently use the terms \emph{processes} and \emph{process identifiers}~(PIDs). Erlang processes communicate with each other by exchanging messages. By knowing the PID of a process, another process is able to send messages to it. Incoming messages are stored in a process mailbox and handled in the order they arrive in.

\section{System Description}
\label{solution}

In this section we describe the OODIDA platform in detail, starting with an overview~(Sect.~\ref{sol:overview}) and execution scenarios~(Sect.~\ref{sol:execution}). The main part is the discussion of the central cloud application~(Sect.~\ref{sol:bridge}). Two brief notes address error handling~(Sect.~\ref{sol:error}) and the backend on client devices~(Sect.~\ref{sol:back}). This section concludes with an overview of practical use cases for our system~(Sect.~\ref{sec:usecases}).

\subsection{Overview}
\label{sol:overview}

\begin{figure}
  \includegraphics[width=\linewidth]{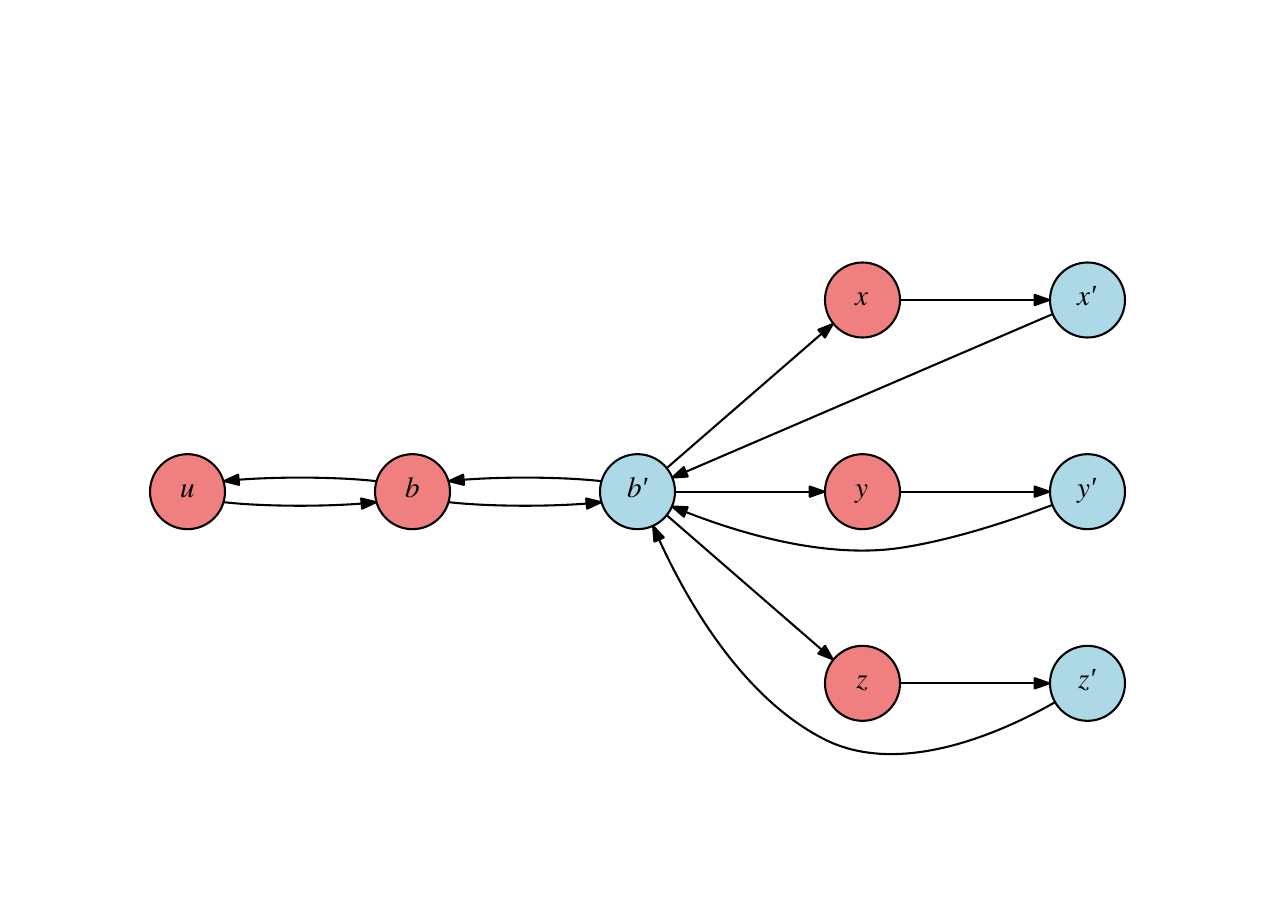}
  \caption{Whole-fleet assignment (one task/client). This execution scenario starts with user node $u$ sending an assignment to cloud node $b$, which spawns an assignment handler $b'$ that divides the assignment into tasks. Red nodes are permanent, while blue nodes are temporary. Each of the client nodes $x, y$ and $z$ spawns a task handler, i.e.\ $x'$, $y'$, and $z'$, that interacts with external applications on the client device.}
  \label{fig:fleet_1}
\end{figure}

\begin{figure}
  \includegraphics[width=\linewidth]{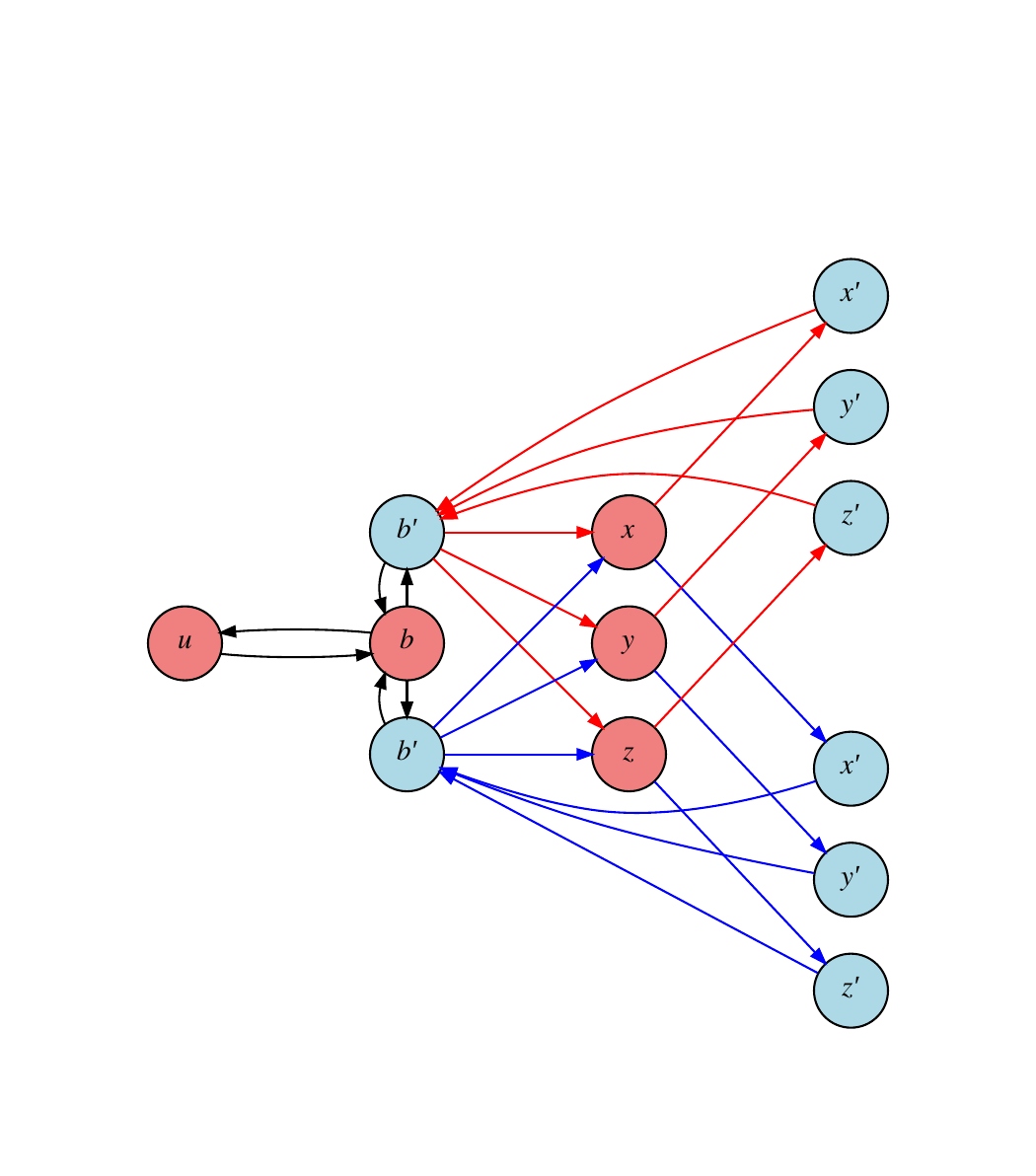}
  \caption{Whole-fleet assignment (two tasks/client). For the base scenario, refer to Fig.~\ref{fig:fleet_1}. Here, we visualize a system configuration in which two assignments are active (cf.~node $b'$). This leads to two task handlers on each client being executed. Note that assignment and tasks handlers are tied to actual assignments. For instance, if node $z$ was not in the set of addressed clients, it would not spawn any task handler.}
  \label{fig:fleet_2}
\end{figure}

OODIDA is a platform for distributed real-time analytics for the automotive domain. Its input is the specification of an \emph{assignment}, that, among others, specifies the \emph{task} to be carried out by the selected subset of clients as well as an optional task for the central cloud server. The task is a specification of the actual computation, while an assignment also contains information on how tasks are to be carried out. Users specify assignments with the help of an external library that facilitates their creation and verification. On client devices, external applications perform analytics tasks. These can be third-party libraries. Our system is able to concurrently execute multiple applications on client devices. The number of users, clients, and tasks is only limited by the computational power of the hardware the system is executed on. Referring to Fig.~\ref{fig:architecture}, the relevant parts for the subsequent presentation of the message-passing infrastructure are the backend application of the user~($u$), the server application~($b$) as well as the assignment handler~($b'$) of the cloud, and the client application~($c$) with its accompanying task handlers~($c'$) on the edge. The sensor manager provides an interface to a third-party application that reads data from the vehicle's CAN bus. The ability to execute user-defined code is hinted at in Fig.~\ref{fig:architecture}, which mentions storage for custom-code. Yet, this feature is discussed elsewhere (cf.~Sect.~\ref{related}).

\subsection{Execution Scenarios}
\label{sol:execution}
In this subsection we describe various usage scenarios. The focus is not on specific assignments, but on their distribution and execution instead. We look at three scenarios: one finite task, multiple finite tasks, and multiple indefinitely long tasks. For simplicity, we consider only one user and three client devices, although there could be arbitrarily many of both. Similarly, our system does not restrict the number of concurrently executed assignments. The used short-hands~($u$, $b$, etc.) for the various components were introduced in Sect.~\ref{sol:overview}. The description is based on whole-fleet assignments. A closing note highlights an important difference for sub-fleet assignments.

% [trim=left bottom right top, clip]

\begin{figure*}
\includegraphics[scale=0.45]{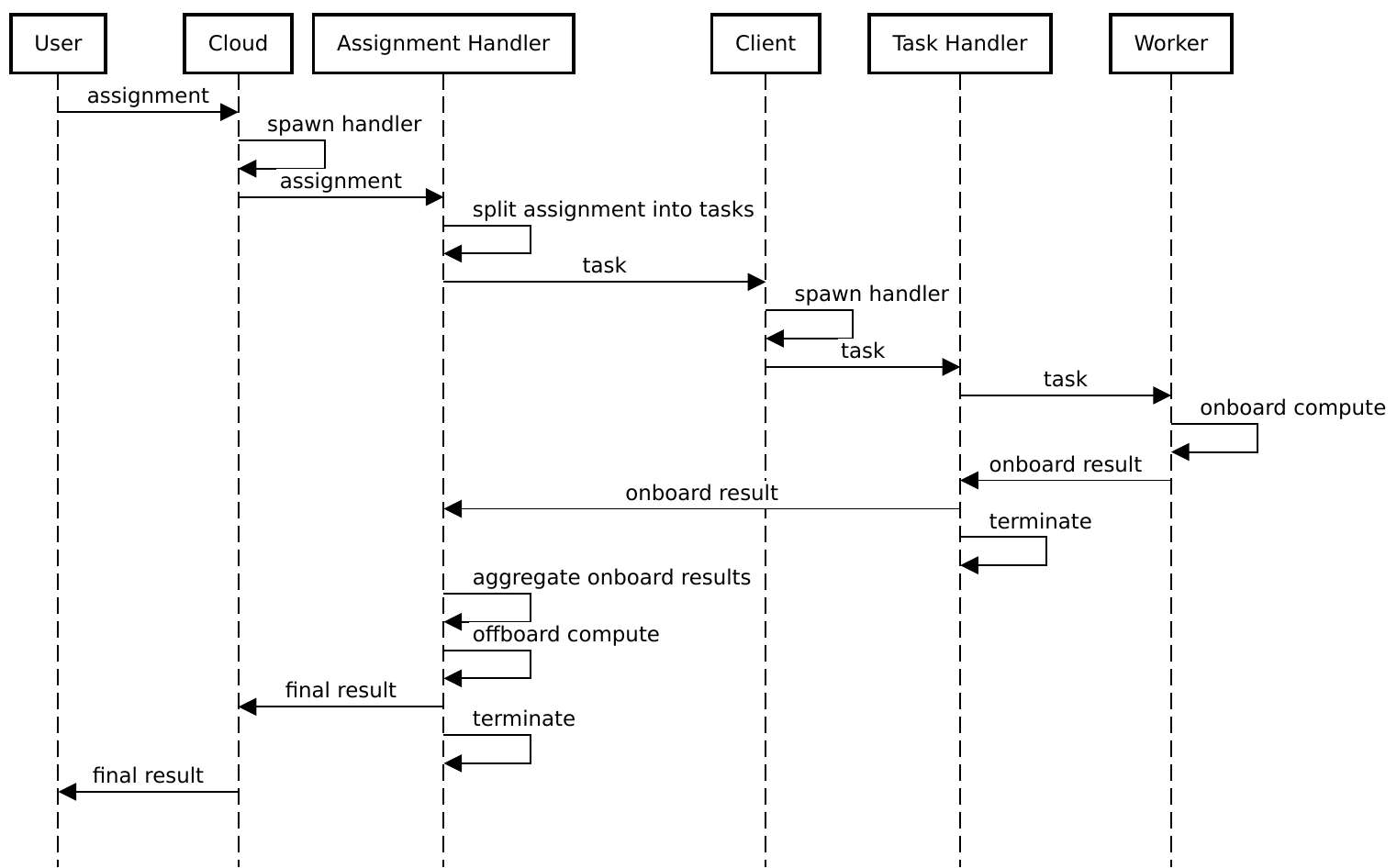}
\caption{Simplified sequence diagram of OODIDA. The system handles multiple users which can all send multiple assignments that can address arbitrary subsets of clients, whereas this diagram only shows one user and one client. On top, clients execute tasks concurrently. The system can also perform iterative work on the task and assignment level. In contrast, this diagram shows an assignment consisting of a task that contains one iteration.}
\label{fig:sequence}
\end{figure*}

The most basic scenario consists of submitting an assignment with \emph{one finite task} for all clients (cf.~Fig.~\ref{fig:fleet_1}): cloud node~$b$ has spawned an assignment handler $b'$, which divides the assignment into tasks and sends a corresponding message to each of the client nodes $x,$ $y$, and $z$. In general, each client node $c_i$ runs on its own OBU. Upon receiving a task description from $b'$, each client process $c_i$ spawns a task handler $c_i'$, which communicates with an external application $a$. Upon completion of the task, the results are sent from $a$ to $b'$ via $c'$. Node $b'$ performs optional off-board computations. Afterwards, results are sent to $b$ and, finally, to the user node~$u$. Executing just one finite task at a time on some or all of the available clients has some practical utility, but many use cases necessitate \emph{concurrent execution of multiple assignments}. For instance, a client that is training a machine learning model with local data could concurrently monitor a range of sensors for critical values for the purpose of performing anomaly detection tasks. A more complicated scenario is shown in Fig.~\ref{fig:fleet_2}, which illustrates a whole-fleet assignment with two concurrent tasks per client. The cloud process has spawned two instances of the assignment handler $b'$, i.e.\ one handler per assignment. Each of the three client processes has likewise spawned one task handler per received task, which means that all client nodes are processing both tasks. Having illustrated these two cases, it is easy to see how additional concurrent assignments are handled. There are also \emph{indefinitely long assignments}, which are a variation of the previous scenarios as tasks could as well run for an unspecified duration. This is the case when execution is not tied to a fixed number of iterations but to a stopping criterion, such as reaching a desired threshold value for the validation error of an artificial neural network. Tasks can also be set up to run until an interrupt from the user has been received. It is also possible to set up an assignment consisting of an indefinite number of iterations. This generates a stream of data that consists of intermediate results, which are produced after each iteration.

When a sub-fleet is addressed, client nodes that do not carry out any task are isolated from the system. OODIDA uses targeted messages that are sent only to affected clients, as opposed to broadcasting to all clients. A task handler is only spawned when there is a task to perform. Furthermore, for each concurrent assignment, a subset can be chosen arbitrarily. It is irrelevant for the system if they overlap or not, albeit there are use cases, such as A/B testing, for which the user should select non-overlapping subsets.

\subsection{Implementation Details}
\label{sol:bridge}
This subsection details the message-passing infrastructure of OODIDA, and how it facilitates scalability with concurrent data analytics workloads. The sequence diagram in Fig.~\ref{fig:sequence} captures the simple case of non-concurrent execution with one user, one client, and one task. In the description below, the fully concurrent case is described, however. The system consists of one central cloud node $b$, $m$ user nodes, and $n$ client nodes that are all intended to communicate via virtual private network. Node $b$ thus provides a connection between the user, i.e.\ a data analyst specifying assignments on one end, and a multitude of OBUs on the other. The underlying message-passing infrastructure has been implemented in Erlang/OTP. Our presentation mainly focuses on this part of the system. We address the external user application only in passing, but provide some details on the client application~(cf.\ Section~\ref{sol:back}). In a production system, each OBU is supposed to execute only one client node. This limit is not enforced, which facilitates an alternative operational mode, i.e.\ simulating a large number of client devices on a more powerful workstation.

\paragraph{Assignment specification.}
Assignments are JSON objects, containing key-value pairs. Among others, they indicate the chosen on-board and off-board computations and the signals to read. The sampling frequency is specified in Hz. Depending on the chosen computations there may be contextual parameters to set. The assignment specification also indicates the number of samples to take, e.g. 1000 samples at 1 Hz, and whether intermediate results are required. The latter is possible by setting a number of iterations. Lastly, we need to choose clients. In addition, there is a keyword for whole-fleet assignments. For sub-fleet assignments, the user can either choose a number of clients or a list of IDs, not PIDs, of clients. 

\paragraph{Server loop.}
On each user workstation, a frontend $f$ is available that interfaces with a Python library. It generates and validates assignment specifications in the JSON format. Once an assignment specification has been generated, it is processed via the user node $u$ and sent via the network to node $b$, which is executed on an internal private cloud. The steps the main server loop $b$ executes are described in Alg.~\ref{alg:node-b}, while the temporary assignment handlers they spawn --- one assignment handler per assignment -- are described in Alg.~\ref{alg:node-bprime}. The main server node $b$ is ready to process incoming assignment specifications or forward results from any of the spawned assignment handlers $b'$ to the user, regardless of the status of any other active assignment because all assignment handlers $b'$ are executed concurrently and independently from each other. Furthermore, $b$ can spawn an arbitrary number of assignment handlers $b'$, which independently await results and process them further. For each new incoming assignment, $b$ spawns a new assignment handler. OODIDA can handle tasks that run for an unspecified amount of time, which can be manifested in multiple ways. For instance, an assignment could consist of one iteration and terminate based on user intervention. Alternatively, it could consist of $k$ rounds, which repeat the main loop in $b'$ $k$ times. However, the duration may be left unspecified, implying that the assignment runs until a predefined stopping criterion has been met or until the user issues a command that stops that assignment. These cases are not shown in the provided pseudocode listings, however.

\algrenewcommand\algorithmicrequire{\textbf{input:}}
\algrenewcommand\algorithmicensure{\textbf{output:}}

\begin{algorithm}[H]
\caption{Central cloud node $b$ (Simplified)}
\label{alg:node-b}
\begin{algorithmic}[1]
\Require Stream of messages $m$, consisting of assignment specifications and assignment results, list of PIDs of available client devices $d$
\While{true}
\If{$m$ is an assignment specification}
\State $p$ := extract assignment parameters from $m$
\State $c$ := extract PIDs of subset of clients from $m$, using $d$
\State $k$ := extract number of task iterations from $m$
\State Spawn assignment handler $b'$ with arguments $p$, $c$, $k$
\EndIf
\If{$m$ is a final assignment result (from $b'$)}
\State Send results to user node $u$
\EndIf
\EndWhile
\end{algorithmic}
\end{algorithm}

\begin{algorithm}[H]
\caption{Assignment handler $b'$ (Simplified)}
\label{alg:node-bprime}
\begin{algorithmic}[1]
\Require assignment parameters $p$, subset of clients $c$, number of rounds $k$
\State $\mathit{acc}$ := $\varnothing$ \Comment Accumulator for results of each round
\For{$i$ in $1 \dots k$}
\State $t$ := task for each client based on $p$
\For{client in $c$}
\State Send task $t$ \Comment Same task for each chosen client
\EndFor
\State $\mathit{acc\_round}$ := $\varnothing$  \Comment Aggregate results of current round
\For{client in $c$} \Comment Simplification; real implementation uses a separate asynchronous
\\  \hspace{10.5em} process that is active until all clients have responded
\State $r$ := completion/results of current round
\State $\mathit{acc\_round}$  := append  $r$ to $\mathit{acc\_round}$ 
\EndFor
\State $o$ := perform optional off-board processing on $\mathit{acc\_round}$
\State $\mathit{acc}$  := append $o$ to $\mathit{acc}$ \Comment Simplification; results of current round may transform\\  \hspace{15em} exiting value $\mathit{acc}$ arbitrarily
\EndFor
\State $\mathit{acc\_final}$ := perform optional off-board processing on $\mathit{acc}$
\State Forward $\mathit{acc\_final}$ to $b$
\end{algorithmic}
\end{algorithm}

\paragraph{User node (incoming).}
Every instance of user node $u$ has two main components, a process \texttt{user} that communicates with node $b$ and a process \texttt{await} that reacts to input sent from the frontend $f$, such as assignment specifications or status requests. We ignore the latter in our presentation, however. When an assignment specification arrives, the process \texttt{await} decodes it, extracts all relevant information, turns it into an Erlang tuple tagged as \texttt{assignment}, and sends it to the process \texttt{user} for further processing. Having a separate process \texttt{await} dedicated to handling incoming assignments enables the process \texttt{user} to remain responsive at all times.

\paragraph{Cloud node (incoming).}
The process \texttt{user} forwards the relevant content of an assignment specification to the process \texttt{cloud} of node $b$. Its state contains the list of currently active clients, pairing their IDs and PIDs.\footnote{IDs are static, while PIDs may change when a client node reconnects. For this reason, the user addresses client nodes with their IDs instead of their PIDs.} The goal is to send task specifications to the chosen client nodes. However, we cannot block the execution of the process \texttt{cloud} until we have received a result because the user may want to issue another assignment in the meantime. This necessitates concurrent execution, which is addressed as follows. The process \texttt{cloud} awaits assignments from the process \texttt{user}. If there is currently no client process available due to no vehicles having been registered in the system, the assignment is dropped (and the user informed of that outcome). Otherwise, the process \texttt{cloud} spawns an assignment handler $b'$ for the current assignment that sends task specifications to the chosen client nodes and monitors task completion. Each incoming assignment leads to spawning a separate node $b'$.

The assignment handler is a key component for enabling concurrent processing of assignments. It receives as arguments the list of currently active clients, the subset of clients~$\boldsymbol{c}$ specified by the assignment, the assignment configuration, and the PID of the process \texttt{cloud}. It first determines the PIDs of the clients specified in the assignment configuration and afterwards sends the specification of the on-board task to each client node. The next step is to block execution for this instance of $b'$ until it has received results from all client nodes (cf. Sect.~\ref{sol:error} for error handling). This is possible due to the cloud process spawning a new assignment handler process for each incoming assignment. We continue at this point after discussing the client node.

\paragraph{Client node (incoming and outgoing).}
The client node $c$ waits for task descriptions to arrive. There is one process \texttt{client} per client node. Whenever this process receives a tuple that is tagged as a \texttt{task}, the task specification is extracted and turned into a JSON object.\footnote{Decoding is not performed by the task handler because some tasks are status requests or special instructions, e.g.\ a shutdown command. These are handled by the client node; to simplify the architecture of the system, the client node processes all incoming JSON objects.} This is followed by spawning a task handler~$c'$, which communicates with an external application $a$ that carries out the specified task. For each new incoming task, a new task handler~$c'$ is spawned. This mirrors spawning a separate assignment handler~$b'$ for each new assignment on the cloud node $b$ and it likewise ensures that the client node remains responsive. The first objective of $c'$ is to forward the task specification to $a$ and await the completion of that task. It is possible to have different external applications process different tasks, but we only consider the case of one external application. Most tasks consist of reading values from a set of sensors for a given interval and performing a computation on it. The result of this computation is sent to $c'$.

An assignment may consist of multiple iterations. However, this is hidden from the client as it treats each iteration in isolation. In other words, the client is both oblivious to whether it performs an assignment with multiple iterations or just one, and also to the iteration it is currently working on. Furthermore, $c'$ does not communicate with $b$ at all. Instead, it directly sends results to $b'$. This reduces the communication and computation overhead of the client node. After sending the result of the task to $b'$, $c'$ terminates.

\paragraph{Cloud node (outgoing).}
Once $b'$ has received the results from all involved task handlers $c_i'$, it performs the specified off-board computation. If the current assignment consists of multiple iterations, $b'$ calls itself recursively. This leads to one more round of sending out task specifications to client nodes~$c_i$ and waiting for the results. On the other hand, if there are no iterations left, the final result is sent to $b$. Afterwards, $b'$ terminates. This is the end of a cascade effect: First, all task handlers~$c_i'$ send the results to $b'$ and terminate themselves. Then, $b'$ performs the off-board computation, sends the results to $b$, and terminates itself as well. Lastly, $b$ sends the results of the assignment to the user node $u$. 

\paragraph{User node (incoming).}
On the user node, the default approach is that a JSON object with the result of the assignment is sent to the frontend application $f$. These results can be further manipulated in $f$, for instance by integrating them into computational workflows where the results of one assignment are used as input for a subsequent assignment (cf. Sect.~\ref{sec:usecases}).

\subsection{Error Handling}
\label{sol:error}
So far, we have made the assumption that all assignments complete and that all client nodes are available all the time. Of course, it is overly optimistic to assume that the network is always stable. To account for real-world disturbances, we use various Erlang/OTP mechanisms for fault-tolerance such as links and monitors. If the connection to a client node has been lost, it is given a set duration of time to reconnect. As long as a disconnected client node is alive, it actively tries to reconnect to the cloud node. Should this fail, its results are excluded from the current iteration of the assignment. For all remaining iterations of the assignment, OODIDA attempts to send a task description also to clients that were lost during previous iterations. This enables them to rejoin this assignment in case they have become available again. The error handling mechanisms are preliminary (cf.~Sect.~\ref{future}).

\subsection{Client Backend}
\label{sol:back}
The external client application, referred to as a \emph{worker} in Fig.~\ref{fig:sequence}, has been implemented in Python, but there is also a functional Go prototype. These applications consume task specifications, perform computational work, and send the results as JSON objects to their task handler. Workers are integrated with a third-party application that provides access to the vehicle's CAN bus data. OODIDA could be run with multiple applications on client devices. An example is delegating stream processing tasks to an external stream processing engine, e.g.\ Apache Edgent.\footnote{The homepage of the Apache Edgent project is \url{https://edgent.apache.org/}.}

\subsection{Practical Use Cases}
\label{sec:usecases}

\begin{figure}
\centering
\includegraphics[scale=0.40]{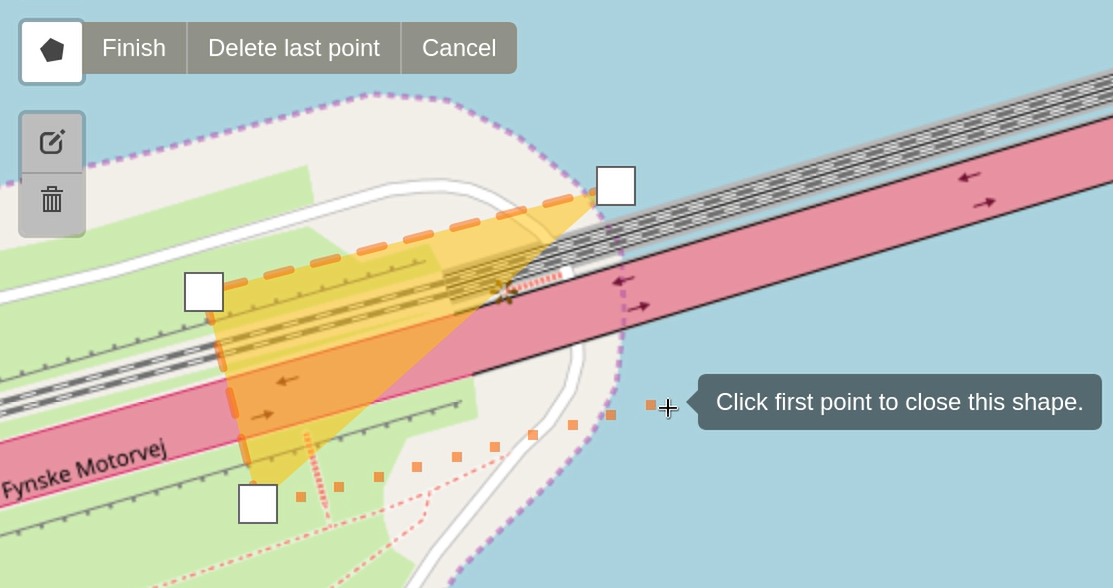}
\caption{Geo-fencing use case screenshot (best viewed in color). This shows how the user frontend facilitates the intuitive definition of geo-fences via defining polygons graphically. A geo-fence works in conjunction with analytics code that processes the detected vehicle coordinates. Geo-fences can be arbitrarily complex, however. In those cases, a definition in code may be preferable.}
\label{fig:geofence}
\end{figure}

OODIDA is a general-purpose system. The examples below are intended to illustrate examples of how this system can be used for practical data analytics tasks. We cover straightforward tasks such as random sampling and anomaly detection, but also more complex ones, such as predictive maintenance, MapReduce-like computations, and federated learning. Lastly, we show how a data analyst can define control flows that use the results of an assignment as the input for a subsequent one. These are all implemented examples, albeit predictive maintenance is a proof-of-concept at this point. 

\paragraph{Sampling.}
One of the challenges of big data is its immense volume. A straightforward approach to handling it is to collect a random sample that amounts to only a fraction of the original data. The assumption is that the random sample has properties similar to the original data. In order to do this with OODIDA, the data analyst needs to set up a sampling task with a list of signals and specify the amount of data in percent that should be collected. This is the on-board task. A fitting corresponding off-board task is the collection of values, paired with the ID of the client they originated from.

\paragraph{Anomaly detection.}
The goal of anomaly detection is to find outliers in data, e.g. determining if a given data point is outside the range a machine learning model predicts. Concretely, assume we have a probability distribution of our data and want to detect values that are considered statistically improbable. Thus, the data analyst selects a model and sets the desired parameters as well as the list of signals to monitor. This defines the on-board task. A fitting off-board task is collecting the device IDs of the targeted OBUs and their corresponding values.

\paragraph{Predictive maintenance.}
Using an existing machine learning model that is tied to a list of signals, it is possible to detect components that may be in need of maintenance. As a simplified example, consider a model for braking pad lifetime that uses braking pad characteristics, the speed of the vehicle, vehicle weight, current precipitation, force of braking and road condition, which is a lookup based on the current GPS coordinates of the vehicle. The on-board task is detecting if braking efficiency deviates from expected values. The corresponding off-board task consists of collecting values and IDs of vehicles that deviate from the model.

\paragraph{Geo-fencing.}
OODIDA has been designed for use with commercial vehicles, which entails several use cases for geo-fencing, such as intrusion detection. Geo-fences can be set up graphically in the user interface (cf.~Fig.~\ref{fig:geofence}) or defined in code. Depending on the complexity of the use case, the computation using a geo-fence could be based on predefined methods or defined as custom code~(cf.~Sect.~\ref{future}). Conceptually, use cases built upon geo-fencing are examples of anomaly detection. Corresponding code runs constantly on all targeted vehicles and when a corresponding event triggers, an alert containing the vehicle ID and relevant additional data is sent from the client to the server and forwarded to the user.

\paragraph{MapReduce-like tasks.}
While running large-scale MapReduce jobs on a fleet of connected vehicles may sound outlandish, there are nonetheless use cases where a MapReduce-like approach is helpful. For instance, consider driver-specific speed-distribution histograms. The corresponding on-board task consists of mapping velocity signal values to a bin value, resulting in $(\mathit{bin}_i, 1)$ and subsequently reducing these pairs to their sums $(\mathit{bin}_i, \mathit{sum})$ as well as normalizing them. Fitting off-board tasks are collecting those histograms or averaging them to one histogram of the entire fleet of vehicles.

\paragraph{Federated learning.}
A much more complex use case is federated learning, which was previously mentioned in Sect.~\ref{p2}. This maps to OODIDA as follows: The assignment handler~$b'$ that is spawned by cloud node~$b$ initializes the global model with arbitrary parameter values and afterwards sends it to the chosen client nodes as a local model for training. After concluding this training pass, each task handler sends its updated local model to~$b'$. Subsequently, $b'$ creates a new global model based on the updated local models and evaluates it with a validation data set. If the resulting error exceeds $\epsilon$, training continues by sending the new global model to clients for training, and so on. Otherwise, training is considered complete and the model is sent to the user.

\paragraph{Control flows.}
Above, we have discussed isolated tasks. However, the user frontend also facilitates control flows, where the results of one assignment serve as the input of a subsequent one. For example, consider the case of predictive maintenance, for which the user issues a whole-fleet assignment. We may start with a model that does not require many resources, does not report false negatives, but has a small risk of reporting false positives. After this assignment has concluded, we filter out the client IDs of all positives, which is likely a fraction of all active client IDs. Then we deploy a computationally more demanding model that has a reduced risk of reporting false positives to those clients. The previous example uses conditional branching. Other control flows may be based on loops that iterate through a sequence of whole-fleet assignments. Such tasks can be set up to concurrently target non-overlapping subsets of clients, which enables A/B testing.

\section{Evaluation}
\label{sec:eval}
The key part of OODIDA is its message-passing infrastructure, which was designed to facilitate concurrent workloads. As the worker applications of our system can be arbitrarily replaced, we do not include experiments on the performance of edge devices as these would be little more than evaluations of the chosen hardware. Furthermore, as many use cases of OODIDA include third-party libraries such as scikit-learn, evaluations based on such use cases would be equivalent to evaluating those libraries, but reveal little of the scalability of OODIDA itself. Indeed, the message-passing infrastructure is taxed regardless of how much computational power has been used on edge devices to arrive at certain results. To give a simple example, the size of a message from the client to the cloud that contains an average of a given number of vehicle speed values is unaffected by the preceding work that was performed on the client to produce that result. A result that was arrived at by taking one value as its input requires as much space, i.e.~a single floating-point number, as one that resulted from averaging a billion input values. Consequently, the scalability and responsiveness of the message-passing infrastructure of OODIDA was evaluated based on tests that stress the ability of a cloud server as well as clients to handle concurrent workloads. The primary focus is on quantifying the ability of the system to remain responsive while concurrently processing assignments. Below, we describe our hardware and software environment (Sect.~\ref{eval:setup}) and the experiments we conducted (Sect.~\ref{eval:experiment}), which is followed by a presentation of the results and a discussion (Sect.~\ref{eval:results}).

\subsection{Hardware and Software Setup}
\label{eval:setup}
We used three quad-core PC workstations as well as Raspberry Pi client devices. Workstation 1 contains an Intel Core i7-6700k CPU (4.0GHz), while workstations 2 and 3 contain an Intel Core i7-7700k CPU (4.2GHz). These workstations are equipped with 32 GB RAM each, of which 23.5 GB were made available to the hosted Linux operating system. They run on Windows 10 Pro (build 1803) and execute OODIDA on Ubuntu Linux~16.04 LTS in VirtualBox 6.0 (workstation 1) or VirtualBox 5.2 (workstations 2 and 3). For the client stress test, we used a Raspberry Pi 3 Model B with a quad-core Broadcom ARMv7 CPU (1.2 GHz) and 1 GB RAM. It runs Raspbian Linux~9 with kernel 4.14.52. This CPU is similar to the CPU of the RVCNs our industrial collaborators use.

\subsection{Experiments}
\label{eval:experiment}
In this subsection we present the experiments we have run to measure the suitability of OODIDA for concurrent workloads. We first describe our considerations behind the choice of those experiments (Sect.~\ref{exp:considerations}) and the data we used~(Sect.~\ref{exp:data}). This is followed by describing the experiments that were executed in order to assess the scalability of client devices (Sect.~\ref{exp:client}) as well as the system, which is based on an evaluation of the server (Sect.~\ref{exp:system}).

\subsubsection{Considerations}
\label{exp:considerations}
OODIDA is designed to run in a heterogeneous system, where a powerful central server orchestrates computationally bound client devices. Our experiments therefore explored the scalability of (a) a single client device and (b) the entire system. For~(a) the mode of connection to the system is irrelevant because the client does not communicate with the cloud process. However, for (b) the difference between Ethernet and wireless connectivity affects the maximal network throughput. Ethernet connectivity is not indicative of real-world usage of OODIDA, but it allows us to estimate the peak performance of our system.

With experiment (a), we seek to determine the scalability, in terms of the number of lightweight concurrent tasks, of our client application executed on resource-constrained hardware. This is important for practical purposes as the hardware we use is very similar to the hardware found in the RVCNs of our project partners. Because those clients are orchestrated by a central server, we use experiment (b) to determine the concurrent workload a standard workstation can handle. Both experiments thus give a good indication of the concurrent workloads that can be carried out. Our experimental setup is idealized, but this is a benefit, not a disadvantage, as execution in a real-world scenario would not be as taxing, considering that in an unstable network, clients can become unavailable, which means that they would not send data that needs to be processed centrally. Furthermore, by simulating a large number of clients in experiment (b), we can generate a workload that is far in excess of what we could achieve in a real-world experiment as we do not have access to a large number of real-world reference vehicles.

One objection, particularly in light of the various use cases we mention elsewhere in this paper, is that we should conduct experiments with those. However, their performance depends significantly on the libraries and hardware chosen. Thus, if we were to do so, we would be measuring the performance of certain algorithms on a particular kind of hardware. Yet, this would not reveal anything about the ability of OODIDA to process a large number of concurrent messages. Furthermore, one may object to the use of virtual machines for executing Linux, which affects the system evaluation, as described below. A non-virtualized environment could be expected to entail an improvement in performance. Given that modern CPUs exhibit hardware virtualization, the overhead of running a VM is relatively modest, however. Thus, the practical implication of the setup of our system evaluation is that native execution would lead to a moderate improvement in performance on the same hardware.  For our use case, the expected performance difference due to native execution is not practically relevant.

\subsubsection{Data}
\label{exp:data}

\paragraph{Client scalability test} As we highlighted earlier (cf.~Sect.~\ref{p1}), using real data can be a tricky situation due to legislation. Yet, for the purpose of evaluating the scalability of the client, we do not need any real data at all. The goal is to concurrently execute a large number of lightweight tasks, which is done via sampling data that is produced by an emulated sensor. One interpretation of this data source would be that it represents a speedometer of a car that is driving relatively steadily. In the end, however, this is not really relevant as the important aspect is that the system is producing floating-point values.
 
 \paragraph{System scalability test} The system scalability test has two aspects that both focus on the responsiveness of the server as the load increases. This is evaluated in two directions: user to server, and clients to server. In both cases, the data used are JSON objects that, in the case of the former, contain an assignment specification, and in the case of the latter, the results of computations. An example of an assignment specification in Python is given in Listing~\ref{assignment}. This is turned into a JSON object that contains the same information in addition to some bookkeeping. The assignment used was very similar to the one shown in the previous listing. The OODIDA user process transforms it into a JSON object that has a size of around 1.2 KB. In order to assess how well the system is able to handle an increasing workload that is generated by clients, we again use synthetic data, as described in the previous paragraph. This time, however, the resulting data is used as the payload in a JSON object that is sent from the clients to the server. This JSON object contains some bookkeeping information and a payload. This is illustrated in Listing~\ref{lst:exch}, which is a simplification of the real-world data that is passed between clients and server. Strictly speaking, the bookkeeping information in the field \texttt{onboard} regarding the computation performed by the client is not necessary, but it simplifies parts of the implementation. This value is a text string. The field \texttt{vals} contains the results of one iteration of one client. For most realistic workloads, this field contains one or more floating-point values in a list data structure. In principle, however, the content could be arbitrary. The payload is normally rather small, often consisting of just one value. An extreme case, and one we consider an outlier, is federated learning (cf.~Sect.~\ref{sec:usecases}), where the payload consists of  matrices. To pick an example that is more demanding than most we have encountered, we send a list containing three floating-point values. This is synthetic data that could be interpreted as the measured odometer value and a lower and upper boundary within which the real value may lie. Again, the point is that the payload consists of floating-point values. What they represent is rather irrelevant for this experiment. This is also obvious in the aforementioned use cases in Sect.~\ref{sec:usecases}, which have the unifying feature that, for the most part, numerical data is sent from the client to the server, where it makes no difference to the system what the data represents. The resulting JSON object has a size of around 0.6 KB.

\begin{figure}
\centering
\vspace{-1.4em}
\begin{lstlisting}[language=Python, caption=Example of an assignment specification, label=assignment]
onboard = Onboard( 
    computation = 'collect',
    signals     = ['speed'],
    filters     = 'x > 100',
    frequency   = 10  # i.e. 10 Hz
    samples     = 3600 * frequency,  # Values of one hour
)

offboard = Offboard(
    computation = 'collect',
    iterations  = 10,  # Run the same assignment 10 times
)

spec = Spec(
    name     = "Sample Assignment",
    clients  = 'all',
    onboard  = onboard,
    offboard = offboard,
)
\end{lstlisting}
\end{figure}

\begin{figure}
\centering
\vspace{-1.4em}
\begin{lstlisting}[language=Python, caption=(Greatly) simplified data exchange format between client and server, label=lst:exch]
{
  "onboard": string
  "vals": numerical data
}
\end{lstlisting}
\end{figure}

\subsubsection{Client devices}
\label{exp:client}
In order to evaluate the scalability of the chosen client hardware, we executed OODIDA with one user. User and cloud nodes were executed on workstation~1, while a Raspberry Pi device executed one client node.  We measured CPU and RAM utilization as the number of concurrently executed lightweight sampling tasks increases. For this experiment, we chose a sampling rate of 1 Hz of input consisting of synthetic CAN bus data. We used Python~3.5.3 and Go 1.11.3. 

\begin{figure}
  \includegraphics[width=\linewidth]{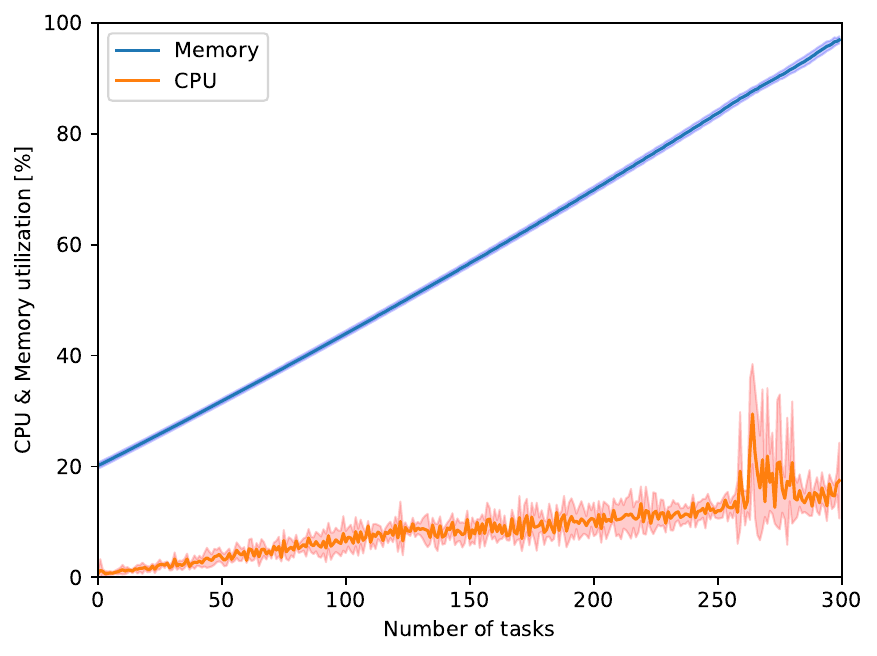}
  \caption{Scalability of the Python client (best viewed in color). This is the result of the concurrency stress test of the Python client application, showing CPU and RAM utilization while increasing the number of concurrently executed sampling tasks at a sampling rate of 1 Hz of a one-dimensional input stream. The plots show the mean (solid color) and standard deviation (shaded area) of five runs. Over 250 tasks can be processed concurrently in Python before hitting hardware limitations. The spike in CPU utilization at $\sim$90\% memory is due to memory swapping. This is due to forks in Python duplicating some memory. At 300 concurrent tasks, CPU utilization is $\sim$18\% with Python.}
  \label{fig:eval_py}
\end{figure}

\begin{figure}
  \includegraphics[width=\linewidth]{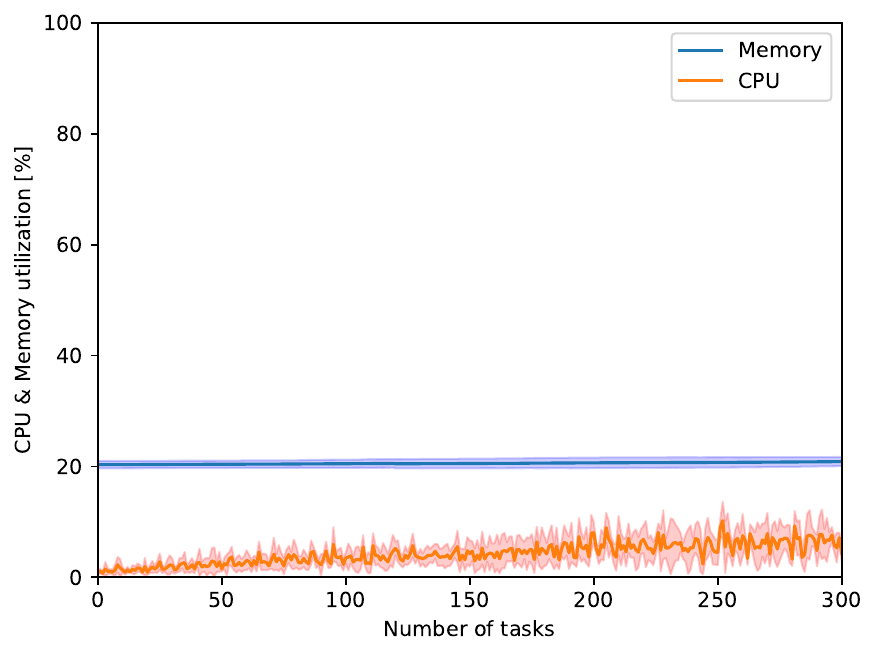}
  \caption{Scalability of the experimental Go client (best viewed in color). The client application was also implemented in Go as a proof-of-concept. See Fig.~\ref{fig:eval_sys_user} for a description of the experiment. In Go, the client uses $\sim$20\% of the available RAM with the same workload. The plots show the mean (solid color) and standard deviation (shaded area) of five runs. At 300 concurrent tasks, CPU utilization is $\sim$10\%. In contrast to Python, it is noteworthy that Go's lightweight \emph{goroutines} only lead to a minimal increase in the amount of memory used as the number of concurrent tasks increases.}
  \label{fig:eval_go}
\end{figure}

\subsubsection{System}
\label{exp:system}
The goal of the whole-system experiment is to determine how OODIDA reacts to surges of activity, with a focus on how large concurrent workloads affect the response time of the cloud process. We looked at this issue from two angles: (a) incoming assignments from the user side as well as (b) incoming results from clients. In (a) the user and cloud nodes were executed on workstation 1. We issued a batch of 100 assignments and measured how the response time, i.e.\ the difference between assignment creation and processing, is affected as subsequent assignments are generated. A surge of such an extent is unrealistically large, so this approach explores the theoretical limits of our system. In (b), workstations 2 and 3 simulated 50 clients each, for a total of 100, for which we issued an assignment with an indefinite number of task iterations. In three different scenarios, they execute either 1, 5, or 10 tasks concurrently per client. We collected 100,000 processing times of a full task iteration per scenario. A full task iteration starts when the assignment handler splits the assignment into tasks for clients and ends when it has collected the results of all clients. The round-trip time between cloud and client workstations is included as well, which sidesteps clock synchronization issues between workstations. In order to put more demand on the cloud node, we implemented dummy client processes in Erlang that produce a result immediately after receiving a task specification, leading to a sustained surge of incoming results. Using 100 clients may seem low, but this is already a multiple of the number of RVCNs OODIDA is targeting. Also, the number of clients is not nearly as relevant as the number of incoming results. Running 10 tasks concurrently on one client is arguably excessive. Yet, the workload of 100 clients running ten tasks is numerically equivalent to, for instance, 500 clients concurrently executing two tasks. Of course, it is also the case that real-world tasks would normally not immediately produce a result. Thus, again, the stress we are putting our system under far exceeds real-world demands.

\subsection{Results and Discussion}
\label{eval:results}
Below, we present the outcome of our experiments. For both client devices (Sect.~\ref{res:client}) and system evaluations (Sect.~\ref{res:system}), the discussion follows right after the results.

\subsubsection{Client devices}
\label{res:client}
We determined RAM and CPU utilization with increasing numbers of concurrently processed assignments, the results of which are shown in Fig.~\ref{fig:eval_py} for Python and Fig.~\ref{fig:eval_go} for a proof-of-concept client application in Go. These plots start with zero assignments, showing that the base utilization of the Python client application is less than 1\% of the CPU and 20.5\% of RAM. With the Go client application, base CPU utilization is likewise less than 1\%, while base RAM utilization amounts to 18.8\%. From that point onward, RAM consumption increases linearly in Python, while CPU utilization fluctuates but mostly increases linearly. Past around 250 assignments, the Python client becomes less responsive. At $\sim$90\% of RAM utilization, CPU utilization spikes due to memory swapping. CPU utilization is close to 20\% at 300 concurrent tasks. In contrast, the Go client application shows a very modest increase of RAM consumption to $\sim$20\%, while the CPU load increased a lot more slowly to $\sim$10\%.

The Python client application exhibits a constant increase of RAM consumption that outpaces CPU utilization. New processes are forks that share some memory. We also tried spawning processes, but this led to an even greater increase in RAM utilization due to not sharing any memory. Another alternative would have been using threads, but we ruled this option out from the start. Despite the drawbacks of forked processes in Python, the number of (simple) assignments that can be executed concurrently is rather high, far exceeding 200. Thus, even devices as resource-constrained as the ones we used are powerful enough for complex real-world scenarios. Again, the issue is not whether the underlying hardware can execute certain tasks, which are highly dependent on particular use cases and hardware capabilities, but whether the OODIDA client can remain responsive as the number of concurrent tasks on an OBU increases. Any bottlenecks can, however, be addressed by horizontal scaling as we could place multiple OBUs on an RVCN. Horizontal scaling may not even be necessary as OBUs use the kind of processors that are also found in smartphones. The performance of these has been improving rapidly and progress in this area shows no signs of slowing down.

\begin{figure}
  \includegraphics[width=\linewidth]{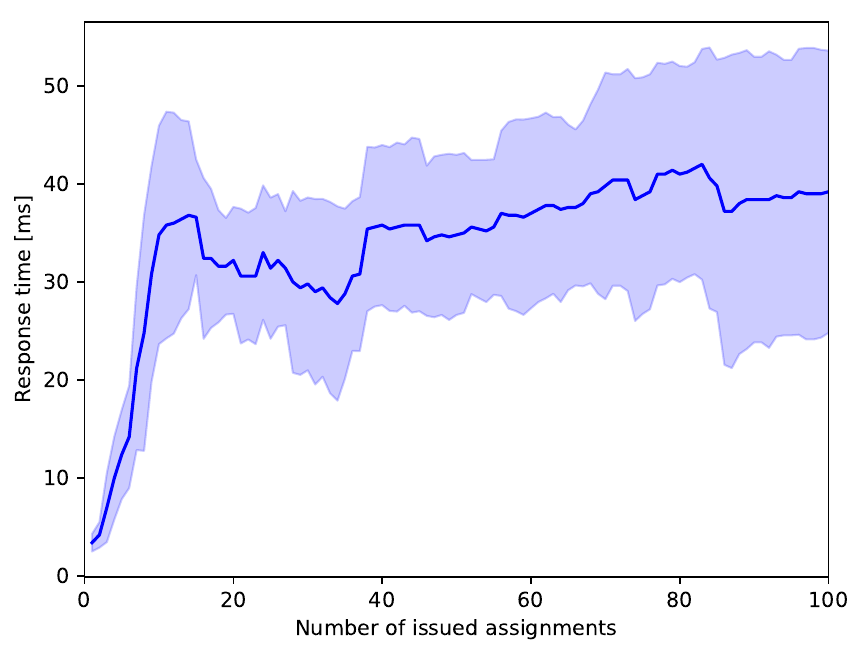}
  \caption{Scalability of the server: Surge processing times of assignments sent by the user (best viewed in color). The stress test of the OODIDA cloud node shows that it can easily handle a surge of incoming assignments. Processing times, based on five runs, which are illustrated by mean (solid color) and standard deviation (shaded area), only marginally deteriorate.}
  \label{fig:eval_sys_user}
\end{figure}

\begin{figure}
  \includegraphics[width=\linewidth]{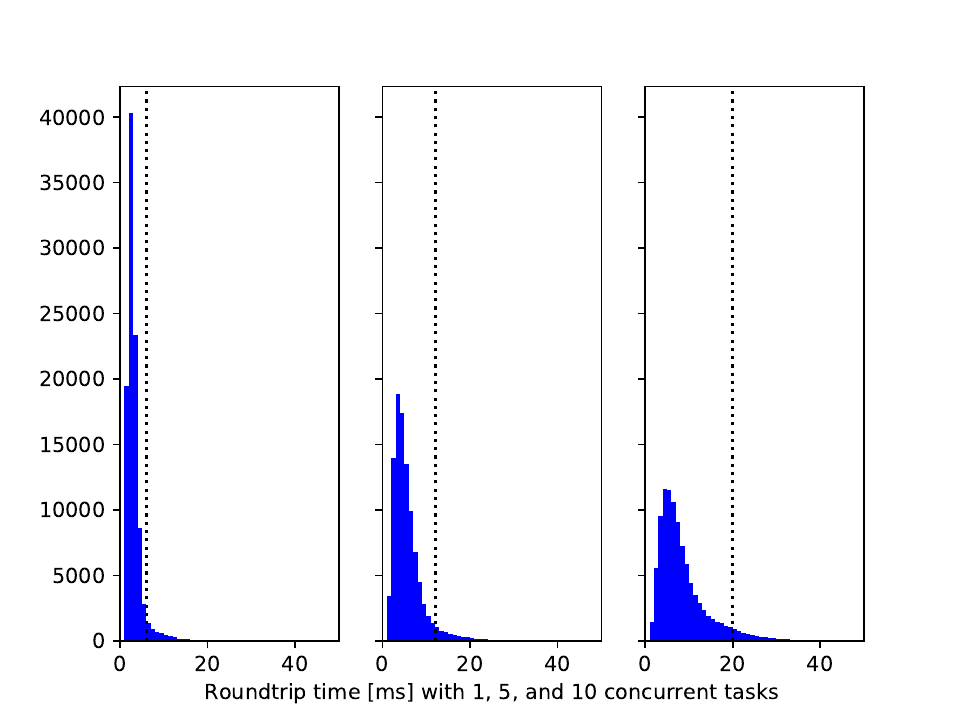}
  \caption{Scalability of the server: Processing times with 95\textsuperscript{th} percentiles (100 clients). This shows that OODIDA comfortably handles large numbers of concurrent assignments. The histograms show the processing times of 100,000 results per scenario, measured from assignment creation on the cloud until the reception of results from the client. We used dummy clients that generated result files directly, i.e.~they did not perform analytics tasks. Because dummy clients can generate many more result files than clients that do perform analytics, these results imply that the OODIDA server application could handle a much higher number of clients in a real-world setting, due to the fact that a much higher number of client devices would be needed to generate the same amount of result files as the dummy clients we used.
}
  \label{fig:eval_sys_clients}
\end{figure}

\subsubsection{System}
\label{res:system}
The results of the evaluation of the system are shown in Fig.~\ref{fig:eval_sys_user} and Fig.~\ref{fig:eval_sys_clients}. The median and standard deviation of five runs in Fig.~\ref{fig:eval_sys_user} indicate that even when submitting an excessively large batch of 100 assignments, the difference between assignment creation by the user and assignment processing on the cloud is less than 40 ms on average, with a slight upward trend. The results of the evaluation of a surge of incoming work from clients and its effect on assignment iteration completion times of the cloud process are shown in Fig.~\ref{fig:eval_sys_clients} and Table~\ref{tab:results}. We see that the system remains highly responsive even as the workload ramps up drastically. In all three scenarios the measurements approximate a log-normal distribution. With 1, 5, and 10 concurrently executed tasks per client and 100 clients, the mean completion time and standard deviation of 100,000 iterations, with median and 95\textsuperscript{th} percentile value in parentheses, are 2.7$\pm$2.1ms (2ms, 6ms), 5.3$\pm$5.0ms (4ms, 12ms), and 8.2$\pm$6.1ms (6ms, 20ms).

\begin{table}[]
\centering{
\caption{Results of the system stress test (cf.\ Fig.~\ref{fig:eval_sys_clients}) with 100 clients performing 1, 5, and 10 concurrent tasks with an indefinite number of iterations. The data are based on 100,000 completed assignment iterations each. All entries are given in milliseconds.}
\label{my-label}
\begin{tabular}{l|rrr}
\hline
\textbf{Concurrent Tasks/Client}  & \textbf{1} & \textbf{5} & \textbf{10} \\ \hline
Mean            & 2.7        & 5.3        & 8.2         \\
Standard deviation            & 2.1        & 5.0        & 6.1         \\
Median          & 2          & 4          & 6           \\
95\textsuperscript{th} percentile & 6          & 12         & 20          \\ \hline
\end{tabular}
\label{tab:results}
}
\end{table}

The main takeaway of our system tests is that OODIDA is able to handle concurrent workloads easily. We have found that congestion due to incoming assignments from users is a non-issue. Even when a batch of 100 assignments arrives at once, which is a figure that is far above real-world requirements, the cloud node keeps up very well. The slight upward trend observed in the data as the number of assignments increases is practically irrelevant with a response time of around 40 ms in the mean. When simulating the workload of an, as of now, unrealistically large number of clients, the system remains highly responsive. As the number of incoming results increases, due a greater concurrent workload on clients, iteration completion times exhibit a modest deterioration. To single out the most extreme case: even with 1000 incoming results per iteration --- 100 clients executing 10 tasks concurrently for 100 iterations --- the 95\textsuperscript{th} percentile is only at 20 ms. Thus, the bottleneck of the system is the performance of client hardware, considering the workload its message-passing infrastructure is able to handle. Furthermore, the central cloud hardware is a regular workstation with a modest number of CPU cores. Thus, we are quite far from reaching hardware limits on that front even if system demands increased by orders of magnitude. We would also like to reiterate that these results hold regardless of the work that is performed on the client as it was excluded from our measurements. We do not foresee scalability of the server to be an issue, given the scope of OODIDA. However, if needed, the server could easily be scaled vertically by using a more powerful machine compared to the standard workstation we have been using. Should that not suffice, the server can be scaled horizontally by adding additional machines and a scheduler that divides the work between them.

\section{Related Work}
\label{related}

When we began developing OODIDA in 2016, there was no mature open-source data analytics framework available that could easily be modified for the demands of the automotive industry. To our knowledge, at the conclusion of this project in early 2020, this still holds. In addition, as would be expected, there is no open-source data analytics framework for analytics in vehicular networks available that could be used with the vehicles of our project partners. In the meantime, potential commercial alternatives such as NVIDIA's EGX or Amazon's AWS IoT Greengrass have been released. However, even if they had been available in 2016, they could not have been used for a variety of reasons, such as vendor lock-in, total cost of ownership, or the problem that some of the data OODIDA processes is too sensitive to be stored on servers owned by a third party.

The problem of using networked devices for data analytics has been tackled from different angles. Google's MapReduce~\cite{dean2008mapreduce} uses clusters of commodity PCs for batch-processing of large-scale finite tasks. An implementation of MapReduce in Erlang/OTP and Python, DISCO, was described by Mundkur et.\ al~\cite{mundkur2011disco}. With MapReduce, a master process keeps track of task completion. While the master may supervise multiple assignments, worker machines are only assigned one specific task along with a corresponding batch of data. In OODIDA, data is generated on clients. They furthermore concurrently execute a multiple of assignments, which could be based on data batches as well as data streams. Apart from those conceptual differences, MapReduce was designed to work with commodity PCs in a data center instead of distributed computationally-bound edge devices. However, due to its inherent flexibility, our system can perform MapReduce-like tasks as well (cf.~Sect.~\ref{sec:usecases}).

As some of our use cases require the ability to process infinite data streams, part of our work could be seen as related to the many stream processing platforms that emerged as a response to limitations of MapReduce. Of course, the focus should not necessarily be on MapReduce, as there are decades of research behind stream processing~\cite{stephens1997survey}. OODIDA can perform complex stream processing tasks, similar to use cases that are tackled in systems like Apache Storm~\cite{toshniwal2014storm} or Apache Spark~\cite{zaharia2010spark}, but with the caveat that our system is intended to orchestrate CPUs of limited computational power instead of data centers housing industry-grade server hardware. However, using our system for stream processing would be rather limiting. While one could write, for instance, an application for filtering values from an infinite data stream, it would be more practical to use an external stream processing engine (SPE) instead, but preferably one that has been designed for that kind of hardware, and interface it with the client node. Due to hardware limitations, using standard SPEs like Spark was infeasible due to their comparatively steep performance requirements. One exception is Apache Edgent, which targets resource-constrained devices. Nonetheless, building a client application on top of Edgent seemed too limiting as our framework has been designed for general-purpose data analytics. The Apache Edgent project has been retired in the meantime, which illustrates one of the risks of wanting to use available off-the-shelf components for internal projects, i.e. the lack of control over development. Furthermore, even if Apache Edgent was still active, it would arguably be difficult if not impossible to get the project maintainers to make changes that accommodate our needs. As a consequence, it would be necessary to maintain a local fork, which would eventually erode the benefits of using an existing solution.

OODIDA emerged from \texttt{ffl-erl}, a framework for federated learning in Erlang~\cite{ulm2019b}. The main conclusion of that project was the general feasibility of using Erlang/OTP for large-scale task distribution. However, it also confirmed that it is not prudent to rely entirely on Erlang/OTP as the performance for numerical computations may not be favorable. In addition, there are practical limitations, such as the limited number of skilled Erlang/OTP programmers on the labor market or the fact that extensive machine learning libraries exist for other programming languages, such as the Python libraries scikit-learn~\cite{pedregosa2011scikit} and Keras~\cite{chollet2015keras}. Both aspects, but especially the latter one, point to the importance of interoperability and the benefits of providing a language-independent JSON interface for external applications. We furthermore have used OODIDA, or derivatives of it, for research in machine learning. This includes a performance evaluation of our own implementations of various federated learning algorithms~\cite{nilsson2018performance} as well as ongoing work on federated learning of neural decision forests~\cite{sjoberg2019federated}.

\section{Future Work}
\label{future}
OODIDA is a fully functional prototype that has been deployed to both stationary OBUs (wired and wireless) and wirelessly connected in-vehicle OBUs. For the former, we put OBUs in a harness, which produce synthetic data or use recorded CAN bus data, and for the latter we used prototype vehicles with OBUs, emitting live data, that wirelessly communicated with a central server. Deploying to a larger population of prototype vehicles, followed by a roll-out to a small fleet of commercial vehicles, is an obvious next step. In addition, in order to provide a more fleshed out solution for the automotive industry, we intend to implement further distributed data analytics algorithms or, when suitable third-party solutions exist, interface with them through the external client application.

We glossed over the practical issue of adjusting the number of tasks a client is allowed to concurrently execute as well as the number of clients per assignment. Both could be addressed via load-balancing. Our current approach consists of using hard-coded values for the number of tasks per client, which means that a client that has reached its maximum number of tasks will be skipped and a different one chosen, provided the user picked arbitrary selection of clients as the distribution mechanism. For the time being, we consider this approach sufficient. However, should OODIDA be used for computationally heavier work, or clients assigned with a higher task load than expected, we would have to reconsider this approach as we would need to more optimally use available resources. Our  ideas in that regard are centered around monitoring RAM and CPU load on the client, and relying on usage metrics for dynamically rejecting tasks, i.e.~if a client has reached a load beyond a predefined threshold, no new tasks will be accepted. It may be necessary to develop more sophisticated approaches that use time-series analysis to predict upcoming spikes in resource usage. In any case, this change could be implemented as an additional step in Alg.~\ref{alg:node-b} that gathers available clients before determining the actual set of clients used for an assignment.

The central server is currently a single point of failure. For a large-scale deployment, this issue has to be addressed by adding redundant servers. Similarly, as the number of client devices grows, so does the expected consumption of network bandwidth. Consequently, by replacing the JSON interface with Protocol Buffers, which use binary data, we may be able to reduce network bandwidth consumption by more than 50\% ~\cite{maeda2012performance}. On a partly related note, network reliability also affects error handling. Our current solution (cf.\ Sect.~\ref{sol:error}) is sensible as long as the network is reasonably stable. As this cannot be generally assumed, we intend to explore alternative approaches. Other work that is still ongoing includes an experimental feature for rapid prototyping, i.e.~the ability of the user to define and execute custom on-board and off-board computations, expressed as Python modules, without having to restart any part of the system~\cite{ulm2019active}. This is relevant for exploratory experiments as it sidesteps the time-consuming and potentially disruptive deployment process.

Security is a potential weakness of Erlang/OTP and an area that deserves paying close attention to~\cite{rodrigues2018towards}. As our system is deployed on a private cloud and executed on a virtual private network, it is less of an issue for us. That being said, we are aware of potential security issues and intend to devote resources to code hardening. We also intend to make further use of the rich Erlang/OTP ecosystem. While we already use Dialyzer~\cite{sagonas2005experience} for static analysis, we intend to also use the type checker Gradualizer,\footnote{Gradualizer is available at: \url{https://github.com/josefs/Gradualizer}.} which is a good match since we use type specification in our entire code base.

\section*{Conflict of interest}
The authors declare that they have no conflict of interest.

\begin{acknowledgements}
This research was financially supported by the project On-board/Off-board Distributed Data Analytics (OODIDA) in the funding program FFI: Strategic Vehicle Research and Innovation (DNR 2016-04260), which is administered by \mbox{VINNOVA}, the Swedish Government Agency for Innovation Systems. It took place in the Fraunhofer Cluster of Excellence "Cognitive Internet Technologies." Vincenzo Gulisano (Chalmers University of Technology), Hans Svensson~(Quviq), and Daniel Lee~(Google) provided insightful comments on a draft of this paper.
\end{acknowledgements}

% BibTeX users please use one of
%\bibliographystyle{spbasic}      % basic style, author-year citations
\bibliographystyle{spmpsci}      % mathematics and physical sciences
%\bibliographystyle{spphys}       % APS-like style for physics
%\bibliography{references}   % name your BibTeX data base

\end{document}